# Inversion of magnetization anisotropy in low magnetic fields for textured Bi2223 superconductor due to magnetic responses from superconducting crystallites and intercrystalline boundaries

D.A. Balaev*, S.V. Semenov, D.M. Gokhfeld, M.I. Petrov

Kirensky Institute of Physics, Federal Research Center “Krasnoyarsk Scientific Center”, Siberian Branch, Russian Academy of Sciences, Krasnoyarsk, 660036 Russia

*E-mail: dabalaev@iph.krasn.ru

A detailed study of the magnetic anisotropy of a bulk textured high-temperature superconductor Bi2223 has been carried out. In the range of moderate and high fields (up to 1 kOe), the difference ΔM between the upper and lower branches of the magnetization hysteresis loop is larger when the external magnetic field H is parallel to the normal of the texturing plane ($H||N_{TP}$) than when H is parallel to the texturing plane ($H\perp N_{TP}$). This magnetic anisotropy is typical for such textured superconductors. However, at small values of the maximum applied field $H_{max}$ (up to 15 Oe at a temperature T = 78 K), an anomalous inversion of anisotropy has been found: $\Delta M(H||N_{TP}) < \Delta M(H\perp N_{TP})$. It has been established that in the field range up to ~30 Oe (the first penetration field into Bi2223 crystallites is about 30 Oe at T = 78 K), the magnetic hysteresis is determined only by the magnetic flux trapped in the intercrystallite boundary subsystem. The pronounced platelet shape of the Bi2223 crystallites leads to partial screening of the external field inside the sample when the field is oriented parallel to the crystallite c-axis. The observed inverse anisotropy $\Delta M(H||N_{TP}) < \Delta M(H\perp N_{TP})$ has been explained by the fact that the effective magnetic field into the intercrystallite boundaries is smaller for the $H||N_{TP}$ orientation than for $H\perp N_{TP}$. Therefore, when the field is oriented $H||N_{TP}$, a larger external field is required for magnetic flux to penetrate into the sample inner than for the $H\perp N_{TP}$ orientation.

## 1. Introduction

Screening of an external magnetic field is a fundamental property of superconductors. It results from the Meissner effect, i.e., the generation of currents in the superconductor that compensate the external field. This property is important for practical applications of superconducting materials in various devices [1,2]. For hydride superconductors, which exhibit superconducting properties at extremely high pressures, experimental observation of external magnetic field screening provided additional evidence for superconductivity at record-high temperatures [3,4].

For type-II superconductors, complete field screening occurs in the range up to the first critical field $H_{c1}$. At higher fields, obviously, only partial field screening takes place. For granular superconductors, the picture of magnetic field screening and its penetration into the superconductor is more complex. In the intercrystallite region superconductivity is suppressed, and the transport of superconducting current through the intergrain boundaries occurs via the Josephson effect. Then complete field screening can occur only up to the first critical field of the intergrain boundary subsystem, $H_{c1J}$, and at $H \geq H_{c1J}$ the field penetrates into the intergrain boundaries in the form of Josephson vortices [5]. In the range from $H_{c1J}$ to $H_{c1G}$, the first penetration field into the superconducting grains, only partial screening of the external field occurs, and Meissner currents circulate mainly inside the superconducting grains.

The picture described above is characteristic of the class of polycrystalline high-temperature superconductors (HTS). For these materials, due to the small value of the coherence length (a few nanometers), a network of Josephson junctions is formed at the developing intercrystallite boundaries. The values of $H_{c1J}$ for $YBa_2Cu_3O_{7-\delta}$ and $(Bi,Pb)_2Sr_2Ca_2Cu_3O_y$ (Bi2223), which are the most popular granular HTS systems with high critical temperatures, are quite small, that are fractions of an Oersted at liquid nitrogen temperature [5]. The values of $H_{c1G}$ for these yttrium and bismuth HTS systems are about 10–30 Oe at liquid nitrogen temperature [6–8].

Thus, in a granular HTS there are two distinct superconducting subsystems: HTS grains and intergrain boundaries. These subsystems differ significantly in the values of $H_{c1J}$, $H_{c1G}$, and critical current density. This leads to a specific shape of the magnetic hysteresis loop of a granular HTS. In low fields, Meissner currents circulate throughout the sample, passing through the grains and the intergranular boundaries. When the external field exceeds the value $H_{C1B}$, magnetic hysteresis associated with the intergranular boundary subsystem appears due to pinning of Josephson vortices. With a further increase in the external field, once it exceeds the value of $H_{C1G}$, hysteresis associated with the HTS grains appears due to pinning of Abrikosov vortices. Such behavior of the hysteresis M(H) dependence of granular HTS is known [6, 9–14]. To observe a crossover from pinning of Josephson vortices to pinning of Abrikosov vortices, it is necessary to carefully measure the M(H) hysteresis loops in the small-field range and at various values of the maximum applied field $H_{max}$. It was recently shown [15–19] that the contribution of the intergrain boundary subsystem to the M(H) hysteresis loop of a granular HTS becomes vanishingly small as $H_{max}$ increases. This is due to the interaction of the two superconducting subsystems: the magnetic moments of the HTS grains together with the external field break the Josephson couplings at the intergrain boundaries. Meissner currents no longer flow through the entire sample, and the contribution to magnetic hysteresis from the intergranular boundary subsystem vanishes. The disappearance of the contribution from the intergranular boundaries occurs not only in sufficiently high fields but also in a zero external field after the application of some external field $H_{max}$ [15–17].

This work considers the mutual influence of the grain and intergrain boundary subsystems in textured granular HTS. Using a textured Bi2223 superconductor as an example, we show that screening of the external field by the platelet-shaped crystallites (hereafter for Bi2223 the term "crystallite" is used instead of "grain") leads to a different character of the magnetic hysteresis anisotropy with respect to the field direction and the orientation of the crystallographic c-axes of the Bi2223 crystallites in the range of small (up to 15 Oe) and large fields.

## 2. Experiment

### *2.1. Preparation method and characterization results of textured Bi2223/Ag sample*

The method for preparing textured Bi2223 was proposed and described earlier [20, 21]. The essence of the method is as follows. At the initial stage, a porous sample with low density (26–40% of the theoretical density of Bi2223 [20]) is obtained. Ultrafine silver may be added at this first stage. In the porous sample, randomly oriented Bi2223 crystallites are formed in the shape of plates with a thickness of 1–2 μm and linear dimensions of about 20–30 μm (see Fig. 1a). The thin part of the plates corresponds to the crystallographic c-axis [20, 21] (Fig. 1a). Next, the porous sample is impregnated with alcohol and subjected to uniaxial compression at a pressure of up to 500 MPa, followed by annealing for 30–50 h at 830°C. The result is a bulk material with a pronounced texture of Bi2223 crystallites (see Fig. 1b), in which the c-axis of the crystallites is oriented predominantly parallel to the normal $\mathbf{N}_{TP}$ to the texturing plane.

In this work, a sample of nominal composition 70 vol.% $Bi_{1.8}Pb_{0.3}Sr_{1.9}Ca_2Cu_3O_6$ + 30 vol.% Ag was studied. The diffraction patterns of the obtained samples contain reflections from the Bi2223 structure and from the bcc structure of silver. The superconducting transition temperature of the studied sample is about 108 K according to magnetic measurements. The transition to the zero-resistance state is observed at about 106 K. The transport current direction was perpendicular to $N_{TP}$. The transport critical current density was 220 A/cm$^2$ at T = 77.4 K.

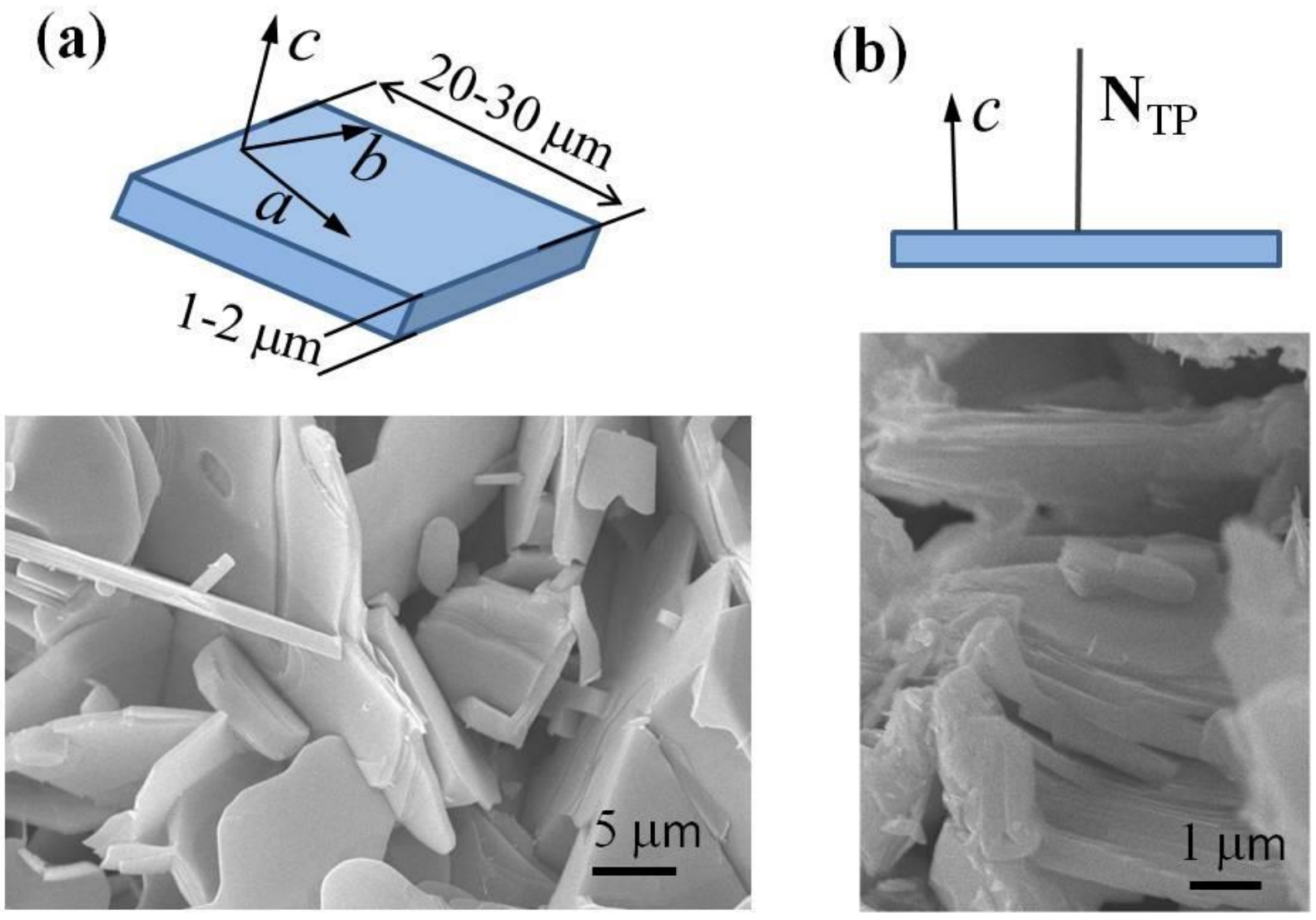

Fig. 1. (a) – Schematic representation of the arrangement of the crystallographic a-b plane and *c*-axis in a platelet Bi2223 crystallite (top) and SEM micrograph of the porous sample (bottom). (b) – Schematic representation of the arrangement of a platelet Bi2223 crystallite (top) relative to the normal $\mathbf{N}_{TP}$ to the texturing plane and SEM micrograph of the textured material (bottom).

*2.2. Magnetic measurements*

For magnetic measurements, the sample was cut into a cube with a side length of about 1 mm. Magnetic measurements were performed on a LakeShore VSM 8604 magnetometer.

The measurements were carried out at a temperature of 78 K for three mutual orientations of the external field **H** and the normal $\mathbf{N}_{TP}$ to the texturing plane: $\theta = \angle(\mathbf{H}, \mathbf{N}_{TP}) = 0°, 45°, 90°$. The magnetic moment of the sample was measured along the direction of the external field. Before each M(H) measurement, the sample was cooled in zero external field. The remanent magnetization $M_R$ was measured at H = 0 after applying a field $H_{max}$. When measuring $M_R$ as a function of $H_{max}$, each subsequent $H_{max}$ value was larger than the previous one (the isothermal remanent magnetization mode was used). The magnetization data are given in units of emu per gram of the whole sample.

### 3. Results and discussion

The M(H) dependences for orientations θ = 90°, 45°, and 0° at maximum applied fields $H_{max}$ equal to ±1000 Oe, ±30 Oe, and ±15 Oe are shown in Figs. 2a, 2b, and 2c, respectively. The asymmetric shape of the magnetic hysteresis loops with respect to the H axis is typical for granular HTS. This asymmetry is caused by weakening of Abrikosov vortex pinning in the near-surface layer of the grains [22].

It can be seen from the data in Fig.2a that the following inequality holds:

$$|M(\mathbf{H} \parallel \mathbf{N}_{TP})| > |M(\mathbf{H}, \theta = 45^0)| > |M(\mathbf{H} \perp \mathbf{N}_{TP})|. \quad (1)$$

For single crystals of Bi2223, a strong magnetization anisotropy is characteristic: the magnetization $|M_c|$ for $\mathbf{H} \parallel c$ can exceed the magnetization $|M_{a\text{-}b}|$ for $\mathbf{H} \parallel a\text{–}b$ by tens of times [8]. In textured bulk polycrystals of, this difference is much smaller [21, 23–28], which is observed for the data in Fig. 2a. This is due to the misorientation of crystallites. Specifically, a deviation of the a-b plane of a crystallite from the texturing plane by only a few degrees, combined with the strict inequality $|M_c| >> |M_{a\text{-}b}|$, leads to the fact that for $\mathbf{H} \parallel \mathbf{N}_{TP}$ the main contribution comes not from $|M_{a\text{-}b}|$ but from the projection of $\mathbf{M}_c$ onto the field direction [21, 27].

Inequality (1) is also valid for the data in Fig. 2b ($H_{max}$ = ±30 Oe), except the vicinity of H = 0, where the remanent magnetization is approximately the same for all three studied orientations of the external field. For $H_{max}$ = 15 Oe (Fig. 2c), a region of fields near H = 0 appears (up to ≈3 Oe, indicated by arrows in Fig. 2c), where instead of inequality (1), the opposite inequality holds:

$$|M(\mathbf{H} \parallel \mathbf{N}_{TP})| < |M(\mathbf{H} \perp \mathbf{N}_{TP})|.$$

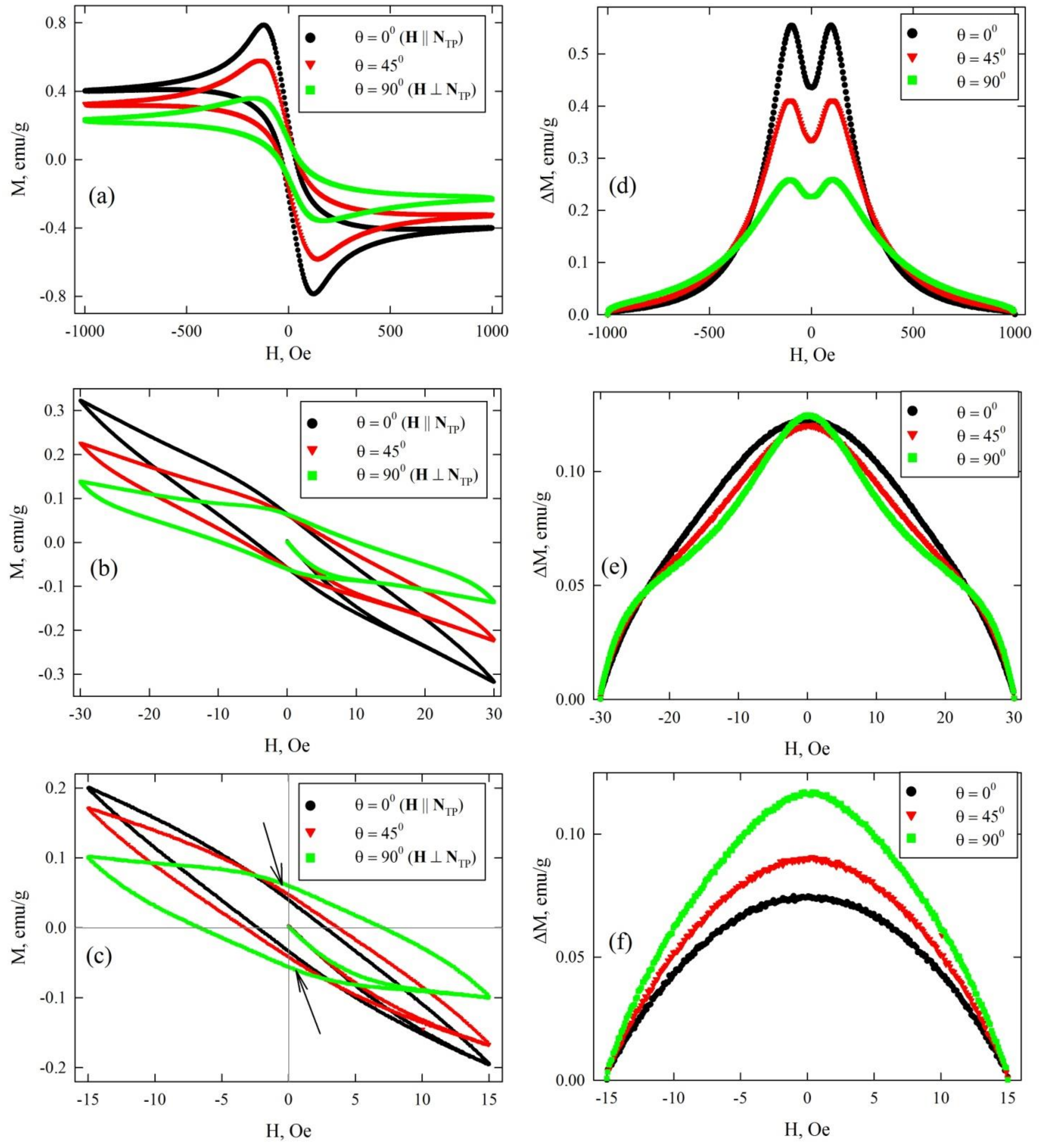

Fig. 2. Magnetic hysteresis loops M(H) for angles $\theta = \angle(\mathbf{H},\mathbf{N}_{TP})$ = 90°, 45°, 0°, measured up to maximum applied fields $H_{max}$ of 1000 Oe (a), 30 Oe (b) and 15 Oe (c) and the respected difference between the hysteresis loop branches ΔM(H) (d), (e), (f).

The unusual behavior of the anisotropic magnetization properties at small $H_{max}$ also manifests itself when considering the difference between the branches of the magnetic hysteresis loop $\Delta M(H) = M_{up}(H) - M_d(H)$, where $M_{up}$ and $M_d$ are the upper and lower branches of the M(H) dependence, respectively. The ΔM(H) dependences for $H_{max}$ values of

±1000 Oe, ±30 Oe, and ±15 Oe are shown in Figs. 2d, 2e, and 2f, respectively. These data show a different character of the anisotropy of ΔM(H) with respect to the relative direction of the external field and the texturing plane of the Bi2223 crystallites for different $H_{max}$ values. For $H_{max}$ = ±1000 Oe (Fig. 2d):

$$\Delta M(\mathbf{H} \parallel \mathbf{N}_{TP}) > \Delta M(\mathbf{H} \perp \mathbf{N}_{TP}), \qquad (2)$$

whereas for $H_{max}$ = ±15 Oe (Fig. 2f):

$$\Delta M(\mathbf{H} \parallel \mathbf{N}_{TP}) < \Delta M(\mathbf{H} \perp \mathbf{N}_{TP}). \qquad (3)$$

Thus, in small fields, an inversion of the anisotropy of the magnetization and its hysteresis occurs with respect to the mutual orientation of the external field and $\mathbf{N}_{TP}$. The observed magnetization anisotropy cannot be caused by the demagnetizing factor of the sample, which is the same for the $\mathbf{H} \parallel \mathbf{N}_{TP}$ and $\mathbf{H} \perp \mathbf{N}_{TP}$ orientations due to the cubic shape of the sample (see section 2.2). To explain this intriguing behavior, it is useful to consider the remanent magnetization $M_R$ as a function of the maximum applied field $H_{max}$ and to estimate the field of magnetic flux penetration into the Bi2223 crystallites.

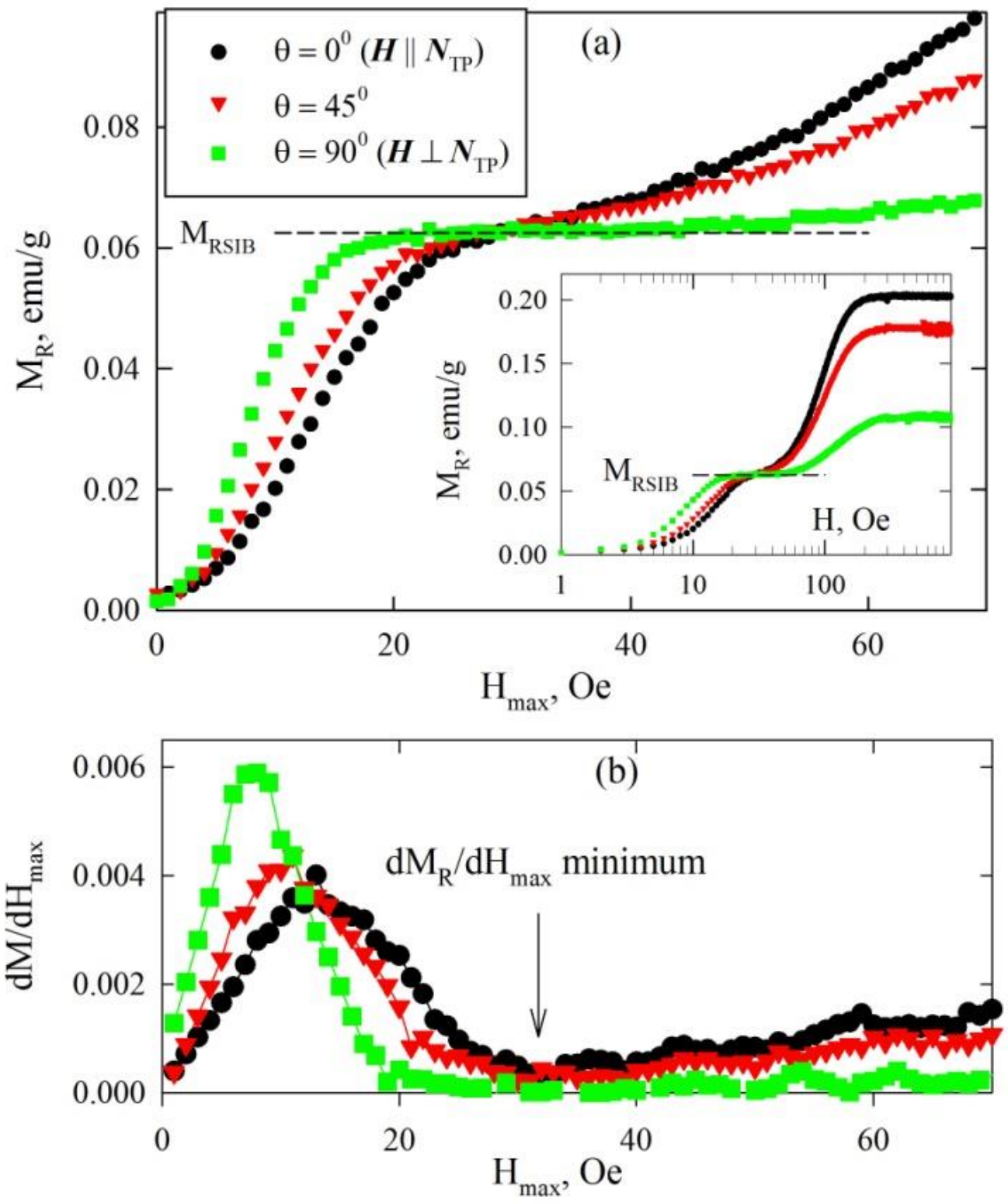

Fig. 3. Dependences of remanent magnetization $M_R$ on maximum applied field $H_{max}$ for angles $\theta = \angle(H,N_{TP})$ = 90°, 45°, 0°, in the range up to 70 Oe (a). Inset shows the same dependencies on a semilogarithmic scale up to 900 Oe. Derivatives $dM_R(H_{max})/dH_{max}$ in the range up to 70 Oe (b).

Figure 3a shows parts of the $M_R(H_{max})$ dependences in the range up to $H_{max}$ = 70 Oe for the three studied orientations of H and $N_{TP}$. The inset of Fig. 3a shows the same dependences in semi-logarithmic coordinates for $H_{max}$ up to 900 Oe. Note that the two-step character of the $M_R(H_{max})$ dependences is a manifestation of the “granularity” of the superconducting material [14–17, 29, 30]. For granular HTS, a similar two-step character is observed in the dependences of resistance on external field [31–34] and resistance on temperature in a magnetic field near the superconducting transition (see, e.g., [35–43]).

The initial growth of the remanent magnetization with increasing $H_{max}$ is caused by flux trapping in the intercrystallite medium. During this stage, the so-called small hysteresis magnetization loop is formed [9–17]. The presence of an intermediate plateau in the $M_R(H_{max})$ dependence in the field range of ~20–30 Oe indicates that the magnetic flux penetration field $H_{c1G}$ for the Bi2223 crystallites significantly exceeds the analogous characteristic ($H_{c1J}$) for the intercrystallite boundary subsystem. In other words, the magnetic hysteresis from the crystallites begins to develop when the hysteresis from the intercrystallite boundaries (the small hysteresis) has already fully developed. The field at which the $M_R(H_{max})$ dependence, after reaching the intermediate plateau, begins to increase again can be reasonably considered close to the value of $H_{c1G}$. Its value can be determined from the condition of the minimum of the derivative $dM_R/dH_{max}$, shown in Fig. 3b. For all three orientations of the external field relative to $\mathbf{N}_{TP}$, the minimum of the derivative is observed at H = 32±5 Oe. Consequently, for all considered orientations of **H** and $\mathbf{N}_{TP}$, the initial segment of the $M_R(H_{max})$ dependence (up to the intermediate plateau) corresponds to the development of the small magnetic hysteresis loop.

From the above, an intermediate conclusion follows: the non-standard character of anisotropy for textured HTS, i.e., inequality (3), occurs for hysteresis associated with processes in the intercrystallite boundary subsystem. Such a non-standard anisotropic behavior is also evident from the data in Fig. 3a: in the range of $H_{max}$ up to $H_{c1G} \approx 30$ Oe, the $M_R$ values for the orientation $\mathbf{H} \parallel \mathbf{N}_{TP}$ are smaller than the $M_R$ values for the orientation $\mathbf{H} \perp \mathbf{N}_{TP}$. For $H_{max} \approx 30$ Oe, the usual behavior is observed: $M_R(\mathbf{H} \parallel \mathbf{N}_{TP}) > M_R(\mathbf{H} \perp \mathbf{N}_{TP})$.

The experimental data, namely the anisotropy of ΔM, indicate that magnetic flux (Josephson vortices) penetrates the sample more strongly for the $\mathbf{H} \perp \mathbf{N}_{TP}$ configuration. Within the critical state model, such behavior could be caused by the fact that when the external field is parallel to long intercrystallite boundaries, magnetic flux pinning is most effective and the intercrystallite critical current density is improved. However, due to the strong anisotropy of the superconducting properties of Bi2223 crystallites, Josephson tunneling across intercrystallite boundaries is preferred along the *a*-*b* crystallographic planes of the crystallites [23–25, 28]. That is, tunneling of carriers between two crystallites along their *c*-axes is unlikely [23, 28]. Therefore, the scenario with different pinning and critical current is questionable.

The influence of the demagnetizing factor *f* of superconducting platelet-shaped crystallites can also be considered [44]. For the case $\mathbf{H} \perp \mathbf{N}_{TP}$, it is small, whereas for $\mathbf{H} \parallel \mathbf{N}_{TP}$ the demagnetizing factor can reach its maximum value. If the magnetization M is negative, then for $\mathbf{H} \parallel \mathbf{N}_{TP}$ the local field $H_{loc}$ acting on an individual Bi2223 crystallite will be larger than the external field H due to the demagnetizing factor (on the return branch of the magnetic hysteresis loop, where the magnetization is positive, $H_{loc}$ will be smaller than the external field [44]). One might expect that for the $\mathbf{H} \parallel \mathbf{N}_{TP}$ orientation the external field would penetrate more strongly into the intercrystallite boundaries, which should lead to an additional increase in $M_R$ and $\Delta M$. However, in the experiment at $H_{max} = \pm 15$ Oe, a decrease in $\Delta M$ is observed for $\mathbf{H} \parallel \mathbf{N}_{TP}$ (3). In sufficiently high fields, the influence of the demagnetizing factor is negligible, and the magnetization anisotropy of the sample is determined by the anisotropic properties of the Bi2223 crystallites. Consequently, the influence of the demagnetizing factor of individual crystallites also does not explain the inversion of the magnetic hysteresis anisotropy.

Further we consider an effect of the interaction of the intercrystallite and crystallite subsystems on the magnetic hysteresis anisotropy. The interaction between the two superconducting subsystems (boundaries and crystallites) occurs as follows. The magnetic moments from the crystallites penetrate through the intercrystallite boundaries, and the corresponding magnetic induction lines are co-directional with the external field, thereby amplifying it (see schematic representations in [15–19, 32–34, 41]). As a result, an effective field the intercrystallite boundaries is determined by

$$B_{eff}(\mathrm{H}) = H + \alpha\, 4\pi M_{CR}(\mathrm{H}). \qquad (4)$$

In this expression, α is a dimensionless parameter characterizing the degree of magnetic flux compression in the intercrystallite medium (the averaged thickness of the intercrystallite boundaries is much smaller than the averaged crystallite size), $M_{CR}$ is the averaged magnetization of crystallites. It has been repeatedly shown, based on measurements of magnetoresistance hysteresis, that the parameter α reaches values of 10–20 [32, 33, 41]. The high value of α leads to the suppression of superconductivity in the intercrystallite boundary subsystem for some value of the maximum field $H_{max}$ = H* (H* ~ 100–300 Oe at 78 K [15–17, 19])]. As a result, the contribution of intercrystallite subsystem to the total magnetization becomes negligible [15–19].

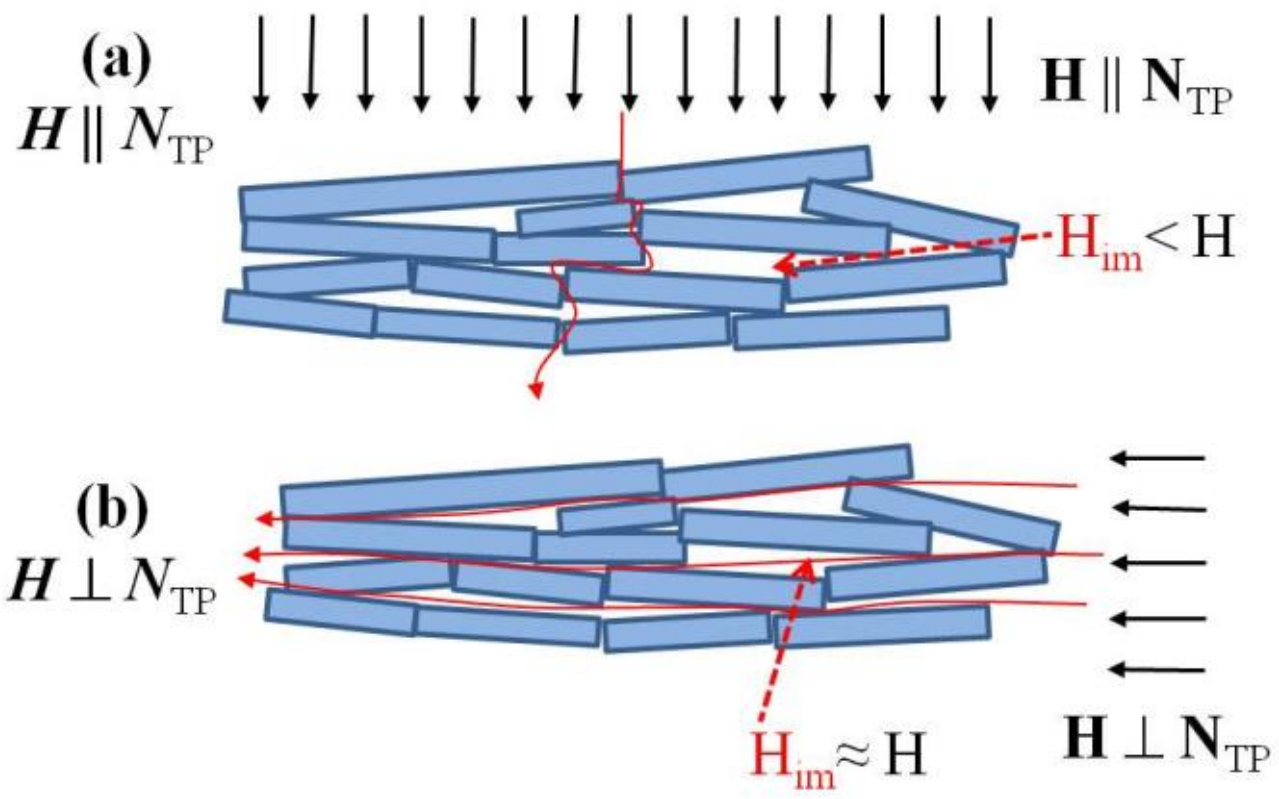

Fig. 4. Schematic representation of the influence of Bi2223 crystallites (filled rectangles) on the field $H_{im}$ inside the textured HTS Bi2223 for the orientations **H** || $\mathbf{N}_{TP}$ (a) and **H**⊥$\mathbf{N}_{TP}$ (b). For **H** || $\mathbf{N}_{TP}$, partial screening of the field occurs in the range up to $H_{c1G}$.

The observed inversion of magnetic hysteresis anisotropy occurs at low fields (up to 15 Oe) smaller than the penetration field of the crystallites $H_{c1G}$ ($H_{c1G}$ ≈ 30 Oe). For the orientation H || $N_{TP}$, the crystallites partially screen the he sample inner, and the external field only penetrate through the intercrystallite boundaries. The surface area of the crystallites is much (by orders of magnitude) larger than the surface area corresponding to the intercrystallite boundaries. Then, an average field $H_{im}$, penetrating in the intercrystallite boundaries, is smaller than the external field. Figure 4a schematically shows this scenario. Figure 4b shows the case when **H**⊥$\mathbf{N}_{TP}$. Here, the surface area corresponding to the thin part of the Bi2223 crystallites does not dominate over the surface area of the intercrystallite boundaries as much as in the case of **H** || $\mathbf{N}_{TP}$. The arrangement of the crystallites favors greater field penetration for the **H**⊥$\mathbf{N}_{TP}$ case compared to **H** || $\mathbf{N}_{TP}$. Based on the above, expression (4) for the effective field in the intercrystallite medium can be rewritten as follows:

$$B_{eff}(\mathrm{H}) = H_{im} + \alpha\, 4\pi M_{CR}(\mathrm{H}_{im}). \qquad (5)$$

Then, for the same external field H, the value of $B_{eff}$ for the $\mathbf{H}\perp\mathbf{N}_{TP}$ orientation, due to partial screening of the external field ($H_{im} < H$), will be larger than for the $\mathbf{H} \parallel \mathbf{N}_{TP}$ orientation. Consequently, the opening of the hysteresis loop in the intercrystallite boundary subsystem with increasing external field (in the range up to $H_{c1G} \approx 30$ Oe) will occur faster for the $\mathbf{H}\perp\mathbf{N}_{TP}$ orientation than for the $\mathbf{H} \parallel \mathbf{N}_{TP}$ orientation. For example, for the magnetic hysteresis loops at $H_{max}$ = 15 Oe, the maximum value of the induction $B_{eff_max}$ ($H_{max}$ = 15 Oe) differs for the $\mathbf{H}\perp\mathbf{N}_{TP}$ and $\mathbf{H} \parallel \mathbf{N}_{TP\ P}$ directions:

$$B_{eff_max}(\mathbf{H}\perp\mathbf{N}_{TP}) > B_{eff_max}(\mathbf{H} \parallel \mathbf{N}_{TP}).$$

As a result, if at H = 15 Oe the loop is not yet fully opened and the $M_R(H_{max})$ dependence has not yet reached the intermediate plateau (Fig. 3a), then for the $\mathbf{H}\perp\mathbf{N}_{TP}$ orientation the loop will be wider than for the $\mathbf{H} \parallel \mathbf{N}_{TP}$ orientation, and inequality (3) will be realized.

At $H_{max}$ = 30 Oe, the $M_R(H_{max})$ dependences reach the intermediate plateau at the same remanent magnetization value $M_R = M_{RSIB}$ (see Fig. 3a). Considering that in the field range up to 30 Oe the contribution to $M_R$ comes mainly only from the intercrystallite boundaries, the same magnetization value $M_{RSIB}$ for the three orientations of **H** and $\mathbf{N}_{TP}$ indicates that the same boundaries contribute to the magnetization for all orientations. It can be assumed that the magnetic hysteresis loop from the intercrystallite boundaries becomes fully developed for $H_{max}$ = 30 Oe. In this case, inequality (3) is no longer realized.

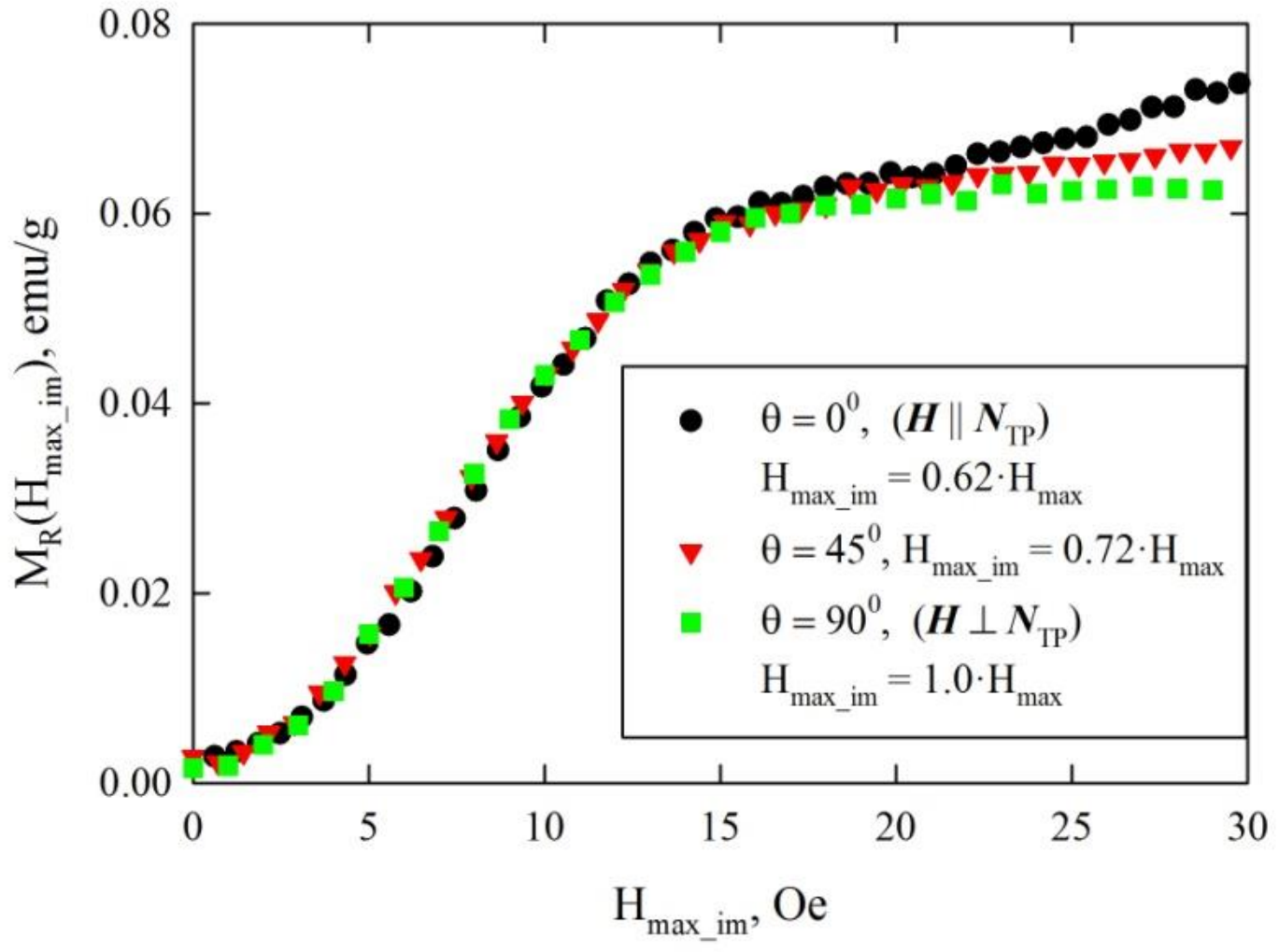

Fig. 5. Scaling of the data from Fig. 3a for the $H_{max}$ range up to 30 Oe. $H_{max_im} = A(\theta)\cdot H_{max}$.

In addition to the above arguments, there are at least two further experimental confirmations of the proposed scenario of field screening manifesting itself in the magnetic properties. If there is only a difference in the value of the field $H_{im}$ in which the boundaries reside for the three orientations (see Figs. 4a, 4b), then scaling of the $M_R(H_{max})$ dependences should be observed when replacing $H_{max}$ with $H_{max_im} = A \cdot H_{max}$. Here it is implied that $H_{im} = A \cdot H$, with the dimensionless coefficient $A < 1$. A is dependent on θ and independent of the external field H. Figure 5 shows the $M_R(H_{max_im})$ dependences for all three orientations. It can be seen that with A = 0.62 for θ = 0°, A = 0.72 for θ = 45°, and A = 1.0 for θ = 90°, the $M_R(H_{max_im})$ dependences agree reasonably well with each other in the range up to $H_{max_im} \approx 18$ Oe. The value $H_{max_im} \approx 18$ Oe corresponds to a field $H_{max} \approx 30$ Oe for the θ = 0° orientation (with A = 0.62). Thus, scaling is observed in the range up to $H_{max} \approx 30$ Oe, in which the magnetic hysteresis is determined by the magnetic response from the intercrystallite boundaries. The obtained values A = 0.62 and A = 0.72 are proportional to the degree of screening of the external field for the $\mathbf{H} \parallel \mathbf{N}_{TP}$ (θ = 0°) and θ = 45° orientations, assuming that for the $\mathbf{H} \perp \mathbf{N}_{TP}$ orientation the field penetrates almost completely.

Another confirmation is the anisotropy of the full penetration field $H_{FP}$. Figure 6a shows parts of the magnetic hysteresis loops at $H_{max} = 30$ Oe for the three orientations of **H** and $\mathbf{N}_{TP}$ in the field range from 0 to +30 Oe. In addition to the lower $M_d(H)$ and upper $M_{up}(H)$ branches of the hysteresis M(H) dependences, Fig. 6a also shows the initial magnetization curves, denoted as $M_{ini}(H)$. The $M_{ini}(H)$ and $M_d(H)$ dependences begin to coincide at $H = H_{FP}$ [43, 45]. To find $H_{FP}$, the differences $\Delta M_{FP}(H)$ between $M_{ini}(H)$ and $M_d(H)$ were plotted (Fig. 6b). In Fig. 6b, $\Delta M_{FP}(H)$ are given in absolute units (emu), and the ordinate axis has a logarithmic scale. It can be seen that, practically for any criterion of $\Delta M_{FP} \approx 0$ to determine the field at which the $M_{ini}(H)$ and $M_d(H)$ dependences begin to coincide, the field $H_{FP}$ will be smallest for the $\mathbf{H} \perp \mathbf{N}_{TP}$ orientation and largest for the $H \parallel N_{TP}$ orientation. Figures 6a and 6b indicate the positions of the $H_{FP}$ fields for the criterion $\Delta M_{FP} = 10^{-5}$ emu (shown as a horizontal line in Fig. 6b). The higher value of $H_{FP}$ demonstrates more effective screening of the external field for $\mathbf{H} \parallel \mathbf{N}_{TP}$.

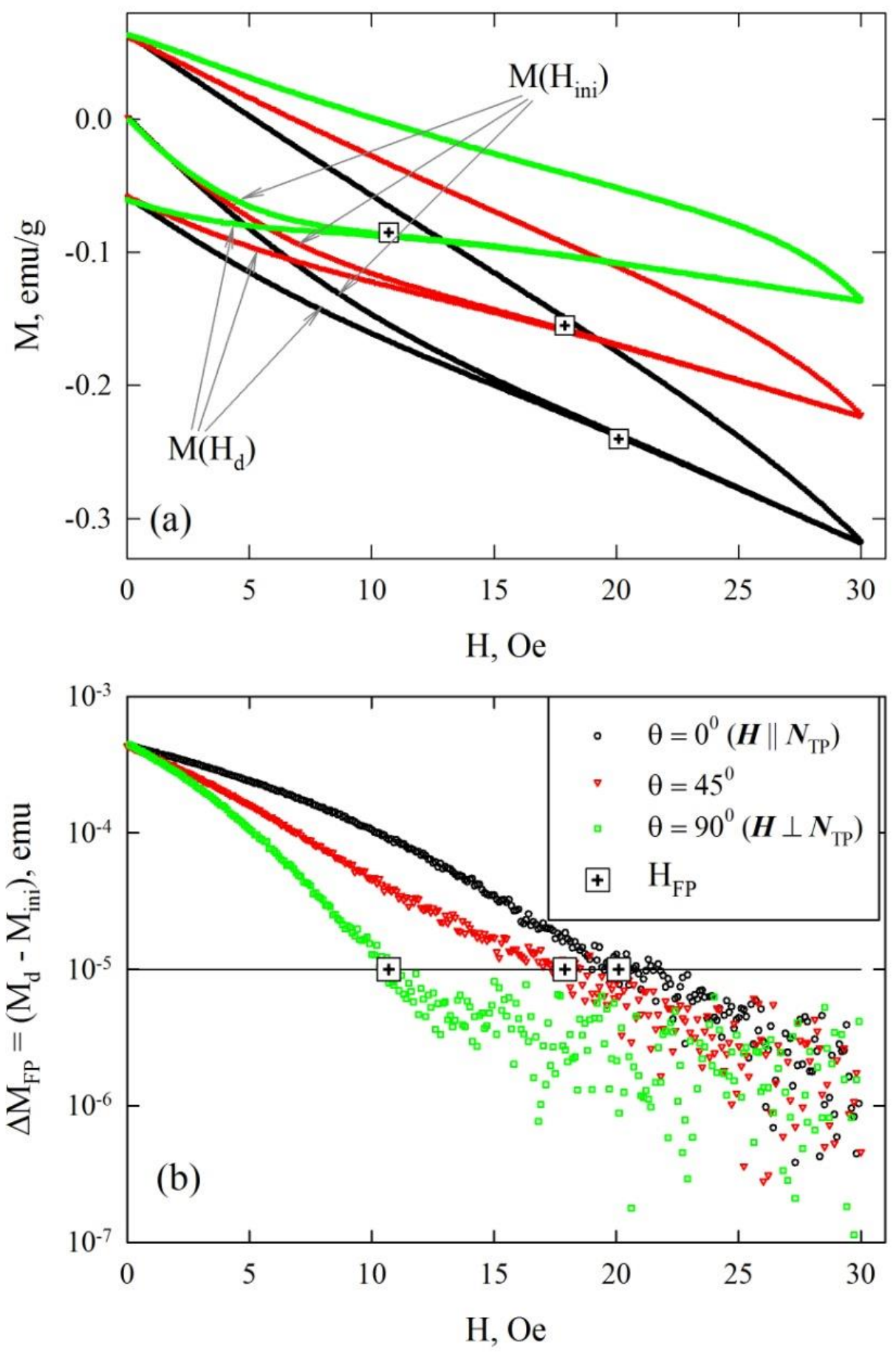

Fig. 6. Segments of M(H) dependences for $H_{max}$ = 30 Oe in the range 0–30 Oe for different orientations of $\mathbf{H}$ and $\mathbf{N}_{TP}$ (a). The branches of the hysteresis loop $M_{ini}(H)$ and $M_d(H)$ are indicated. Difference $\Delta M_{FP}(H) = M_{ini}(H) - M_d(H)$ as a function of external field (b); ordinate axis has logarithmic scale, magnetization data are given in absolute units. In (a) and (b), the points $H_{FP}$ are marked, at which the $M_{ini}(H)$ and $M_d(H)$ dependences coincide for the criterion $\Delta M_{FP}(H) = 10^{-5}$ emu (horizontal line in (b)).

Thus, the experimental facts presented above confirm the realization of the proposed mechanism for the discovered intriguing effect: inversion in anisotropy of the hysteresis loop parameter ΔM (inequalities (2) and (3)). Partial screening of the external field by the Bi2223 crystallites for the $\mathbf{H} \parallel \mathbf{N}_{TP}$ orientation leads to a smaller response from the intercrystallite boundary subsystem in the small-field region.

**4. Concluding remarks**

It has been known for quite some time that the external field begins to penetrate a bulk granular superconductor almost immediately through the intergrain boundary subsystem [5] in the form of Josephson vortices. When cycling the external field to small values, the magnetic hysteresis is determined only by processes in the intergrain boundary subsystem. The superconducting grain subsystem gives a linear and hysteresis-free field response up to the first penetration field into the grains $H_{c1G}$ (for the studied textured HTS Bi2223, $H_{c1G} \approx 30$ Oe at nitrogen temperature). As the field increases ($H > H_{c1G}$), hysteresis associated with the dynamics of Abrikosov vortices in the grains develops, and the magnetic hysteresis loop has contributions from both superconducting subsystems – grains and intergrain boundaries.

The results obtained in this work complement the described picture of external field penetration into a granular superconductor for the case of a textured polycrystalline HTS of the bismuth system with a pronounced platelet shape of the crystallites and anisotropy of the superconducting properties of the Bi2223 crystallites. In moderate and strong magnetic fields, the magnetic response and the distance between the branches of the magnetization loop $\Delta M$ are largest for the orientation in which the field is perpendicular to the texturing plane. This is related to the anisotropy of the superconducting properties of the Bi2223 crystallites themselves. However, in small fields (up to ~15 Oe at nitrogen temperature, for the studied textured HTS Bi2223), the opposite picture is observed: the distance between the branches of the magnetization loop $\Delta M$ is largest for the orientation in which the field is parallel to the texturing plane. This is associated with screening of the external field by the Bi2223 crystallites, which have a plate-like shape. Consequently, in small fields, the distance between the branches of the magnetization loop $\Delta M$, determined only by the response from the intercrystallite boundary subsystem, becomes larger for the orientation of the external field parallel to the texturing plane, where the external field is practically not screened.

**Acknowledgements**

The magnetic measurements were carried out on the equipment of Krasnoyarsk Regional Center of Research Equipment of Federal Research Center «Krasnoyarsk Science Center SB RAS».

This study was carried out within the State assignment of the Ministry of Science and Higher Education of the Russian Federation for the Kirensky Institute of Physics, Siberian Branch of the Russian Academy of Sciences.